\documentclass[preprint,12pt]{elsarticle}
\biboptions{sort&compress}

\usepackage{amssymb}
\usepackage{amsmath}

\usepackage{siunitx}
\usepackage{upgreek}
\usepackage{booktabs}
\usepackage{tabularx}
\usepackage{array}
\usepackage{float}
\usepackage[normalem]{ulem} 

\usepackage{xcolor}

\usepackage{soul}
\soulregister\cite7
\soulregister\ref7
\soulregister\emph7

\journal{Acta Materialia}

\begin{document}

\begin{frontmatter}



\title{Precipitate-Induced Dynamic Strain Aging and Its Effect on the Strain Rate Sensitivity of Precipitation Hardened Aluminum Alloys}

\author[label1]{Sahar Choukir}
\affiliation[label1]{organization={Cornell Fracture Group, Civil and Environmental Engineering},
            addressline={Cornell University},
            city={Ithaca},
            postcode={14850},
            state={New York},
            country={United States of America}}

\author[label1,label2]{Derek Warner}
\affiliation[label2]{organization={Canterbury Fracture Group, Department of Civil and Environmental Engineering},
            addressline={University of Canterbury},
            city={Christchurch,},
            postcode={8140},
            country={New Zealand}}




\begin{abstract}
We examine precipitate-induced dynamic strain aging in precipitation-hardened Al–Cu alloys by combining atomistic simulations, kinetic Monte Carlo, and analytical rate theory. Atomistic simulations were used to characterize (1) the energetics of nearest neighbour Cu$\leftrightarrow$Al exchanges at dislocation - precipitate junctions and (2) the subsequent change in obstacle strength. For robustness, the simulations were performed with two distinct interatomic potentials. The resulting catalog of local Cu–Al exchange events was used as input for a kinetic Monte Carlo model of the time-dependent evolution of obstacle strength during dislocation pinning at the precipitate. The predicted strengthening kinetics were then embedded in an analytical dynamic strain aging model to predict the strain-rate sensitivity parameter.  
On the whole, the modeling predicts a low strain-rate sensitivity across a broad range of intermediate quasi-static strain rates, consistent with experimental observations for precipitate-strengthened alloys. The results therefore identify a mechanistic origin of the low strain-rate sensitivity in precipitation hardened aluminum alloys, emerging directly from the kinetics of dislocation–precipitate interactions when nearest neighbour Cu$\leftrightarrow$Al exchanges are considered.  
\end{abstract}






\end{frontmatter}



\section{Introduction}
\label{Intro}

The dependence of flow strength on loading rate is commonly characterized by the strain rate sensitivity parameter, \(m=\left(\partial \ln \sigma/\partial \ln \dot{\varepsilon}\right)_{\varepsilon,T}\)~\cite{metallurgy1986ge}. The parameter $m$ represents a constitutive characteristic that competes against strain localization and as such can have significant technological implications. Specifically, its value can control elongation ~\cite{hutchinson1977influence,backofen1967knoop}, shear banding~\cite{rudnicki1975conditions}, fracture toughness~\cite{irwin1964crack}, and surface finish ~\cite{metallurgy1986ge,hosford2011metal}. 

Across a broad range of conditions, $m$ is controlled by the thermally activated motion of dislocations in metals ~\cite{kocks1975thermodynamics}. 
In many cases of technological importance, the thermal activation process can be complicated by an energy barrier that evolves with time. The evolution of the energy barrier has long been attributed to the diffusive redistribution of alloying elements near dislocations~\cite{cottrell1949dislocation}.
When the redistribution process occurs on the same timescale as the dislocation motion, the process is referred to as dynamic strain aging~\cite{mccormigk1972model,mulford1979new}, which has traditionally been viewed as the origin of the Portevin–Le Chatelier effect~\cite{le1909influence,portevin1923phenomene}. 

The study of dynamic strain aging has been primarily focused on the diffusive rearrangement of dispersed solute atoms at mobile and forest dislocations~\cite{cottrell1949dislocation,van1982strain,mccormick1988theory,kubin1991dynamic,mccormick1995numerical,picu2004atomistic,soare2008single,zhang2026interplay}. Nonetheless, the concept of dynamic strain aging is not limited to the diffusive redistribution of dispersed solute atoms, as it might also include the diffusive redistribution of clustered alloy atoms, e.g. precipitates. 

It is well established that alloying atoms can be diffusely redistributed by precipitate-dislocation interactions~\cite{lysenko1968change,picu2004atomistic,legros2005pipe,legros2008observation,hutchinson2014quantitative,zhang2021solute,zhang2026interplay} However, the mechanical implications of these interactions are complex and remain to be fully understood. Here, we utilize atomistic-based materials modeling to isolate and study precipitate-induced dynamic strain aging involving only single atom nearest neighbor exchanges. 

The broad consensus is that precipitation (1) reduces the concentration of alloy atoms in solution, (2) subsequently reduces solute induced dynamic strain aging, and (3) ultimately leads to an increase in $m$~\cite{mccormigk1972model,pink1987guinier,kubin1990evolution,jiang2006spatiotemporal,zhang2022critical}. One well-known exception to this trend is Al-Li alloys that have ordered precipitates and low solute concentration. In this case, diffusive alloy atom redistribution is thought to restore sheared precipitates to their low energy ordered state, providing an increase in strengthening at long time scales that acts to reduce $m$ values~\cite{behnood1989plastic,brechet1996pseudo,zhu1998strain,ovri2015new,rowlands2023portevin}. More recently, dislocation induced dynamic precipitation and dissolution of precipitates has been linked to changes in $m$ in Al-Mg-Zn alloys \cite{zhang2026interplay}.

The above examples involve either ordered precipitates or diffusive transport of alloy elements at the scale of the precipitate spacing.  
Here, we present atomistic simulation results for an Al–Cu alloy that show local nearest neighbor exchanges of alloy elements at dislocation-precipitate junctions can lead to dynamic strain aging, i.e. with no transport of dispersed solute atoms or ordered precipitates.    
This result suggests that dynamic strain aging may be more common than previously thought and provides broad reconciliation for the experimentally measured low rate sensitivity of precipitate-strengthened aluminum alloys, $m\approx0.005$~\cite{byrne1961precipitate,muraishi2002strengthening,khan2012variable,gupta2017strain}, with the substantially larger values, 
$m\approx0.04$ obtained from atomistic and continuum models of dislocation–precipitate interactions~\cite{xu2007thermally,singh2013atomistic,saroukhani2016harnessing}. While previous work has shown that this discrepancy may be resolved by accounting for heterogeneous precipitate populations and avalanche-like dislocation behavior at larger length scales~\cite{picu2009strain, mcdowell2024nonequilibrium1,mcdowell2024nonequilibrium2,mcdowell2024nonequilibrium3,mcdowell2025nonequilibrium4}; our results show that there is resolution at the level of a single dislocation interacting with a single precipitate, when diffusive local redistribution of precipitate alloy atoms is included. 

\section{Methods}

This study begins with the description of atomistic simulations that quantify the thermodynamic basis for individual Cu$\leftrightarrow$Al exchanges and the associated strengthening increments. These quantities form the input catalogue for a kinetic Monte Carlo (kMC) model, which evolves the alloy atom configuration in time and yields an activation energy barrier that increases in time, i.e. a strengthening.

A continuum aging law is then fit to the kMC model to extract the characteristic parameters describing the saturation strength and aging timescale. Finally, the continuum aging law is used in the Soare–Curtin rate-theory framework, which models a single aging mechanism: a dislocation pinned at an obstacle whose local chemical environment evolves in time, producing a time dependent escape barrier~\cite{soare2008single}. This connects aging kinetics obtained from atomistic simulation to the strain-rate dependence of the critical resolved shear stress associated with dislocation–precipitate interaction. This section presents each component of this workflow in detail, beginning with the atomistic model and progressing through the kMC simulations to the rate-theory dynamic strain aging analysis based on the analytical Soare–Curtin formulation~\cite{soare2008single}. 

\subsection{Models}
\subsubsection{Atomistic Simulations}
\subsubsection*{Geometries}
Atomistic simulations of edge dislocation interacting with a GP-zone in aluminum were performed using the \textsc{LAMMPS} package~\cite{plimpton1995fast}. The simulation setup reproduces the geometry used in Singh~et~al.~\cite{singh2010mechanisms}, consisting of an fcc Al crystal bounded by the \((\bar{1}10)\), (111), and \((11\bar{2})\) planes along the \(X\), \(Y\), and \(Z\) directions, respectively. Periodic boundary conditions were applied along \(X\) and \(Z\), while the \(Y\)-surfaces were traction-controlled to impose shear. A single edge dislocation with Burgers vector \(\mathbf{b} = \tfrac{1}{2}[\bar{1}10]\) was introduced using the elastic displacement field, with the line direction parallel to \(Z\).

Two GP-zone orientations were considered: one lying on the (001) plane, parallel to the primary slip plane (hereafter referred to as the GP~\SI{0}{\degree} orientation), and another on the (100) plane, which is inclined by approximately \SI{60}{\degree} to the Burgers vector (the GP~\SI{60}{\degree} orientation). The GP-zone was modeled as a monolayer of Cu atoms embedded in an Al matrix. The simulation cell dimensions (\SI{34}{\nano\metre}~$\times$~\SI{42}{\nano\metre}~$\times$~\SI{16}{\nano\metre}, $\sim\!1.4\times10^{6}$ atoms) were selected following the convergence studies of Singh~et~al.~\cite{singh2010mechanisms}, who demonstrated that this cell size yields negligible image and boundary effects for dislocation–precipitate interactions in Al--Cu systems, apart from those expected along the dislocation line direction.

\subsubsection*{Interatomic Potentials}

Given the complexity of dislocation–GP zone interactions, two distinct Al–Cu interatomic potentials were employed to gain a more robust indication of the true behavior. Using both the NNP and ADP creates a complementary framework for studying GP-zones. Their agreement validates the observed trends, while their differences help identify and isolate potential-specific artifacts.

\paragraph{(i) Angular-Dependent Potential (ADP)} The first potential is the Al--Cu Angular-Dependent potential developed by Singh and Warner~\cite{singh2010mechanisms}, which is an angular dependent extension of the Embedded Atom Method (EAM)~\cite{mishin2005phase}. The total energy is given by
\begin{equation}
E_{\text{tot}}^{\text{ADP}} = 
\tfrac{1}{2}\sum_{i\neq j} \upphi_{ij}(r_{ij}) 
+ \sum_i F_{i}(\tilde\rho_i)
+ \tfrac{1}{2}\sum_{i,\alpha}(\mu_i^{\alpha})^2
+ \tfrac{1}{2}\sum_{i,\alpha,\beta}(\lambda_i^{\alpha\beta})^2
- \tfrac{1}{6}\sum_i v_i^2,
\end{equation}
where indices \(i\) and \(j\) refer to atoms and superscripts \(\alpha,\beta=1,2,3\) to cartesian directions. \(\phi_{ij}\) is the pair-interaction potential depending only on the scalar distance \(r_{ij}\) between atom \(i\) and atom \(j\), and \(F_i\) is the embedding energy of atom \(i\) in the host electron density \(\bar{\rho}_i\) induced at site \(i\) by all other atoms of the system. The additional dipole vectors (\(\mu_i\)), quadrupole tensors (\(\lambda_i\)), and scalar (\(v_i\)) terms introduce angular dependence through short-range functions. Complete details of these functions are given in Apostol and Mishin~\cite{apostol2011interatomic}. 

The parameters of the cross-interaction functions ($\upphi_{ij}(r_{ij})$) were fitted, using a database that combined experimental and first-principles (density functional theory, DFT) data for (a) formation energies and elastic constants of the $\theta$ and $\theta'$ precipitate phases, (b) heats of solution and solute–vacancy binding energies for dilute Cu in Al and Al in Cu, (c) formation energies of several ordered Al–Cu compounds spanning Al-rich to Cu-rich compositions, and (d) generalized stacking-fault and shearing energetics for Cu monolayers embedded in Al in geometries reminiscent of idealized GP-zones ~\cite{apostol2011interatomic}. The loss function is the weighted mean-squared error between calculated and target properties. This formalism captures directional bonding effects that are absent in conventional EAM models. All pure-element functions are taken directly from the well-established Mishin EAM potentials for pure Al and Cu~\cite{mishin1999interatomic,mishin2001structural}.

\paragraph{(ii) Neural Network Potential}
The second potential is a neural-network potential (NNP) trained on DFT data for Al--Cu systems~\cite{marchand2020machine}, following the Behler--Parrinello framework~\cite{behler2007generalized} as implemented in the open-source \texttt{n2p2} library~\cite{singraber2019parallel}. The total energy of a structure is expressed as a sum of atomic contributions,
\begin{equation}
E_{\mathrm{tot}}^{\mathrm{NNP}} = \sum_i E_i^{\mathrm{NNP}},
\end{equation}
where each atomic energy $E_i^{\mathrm{NNP}}$ is represented by a feed-forward neural network that maps a set of atom-centered symmetry functions $G_m$ (radial and angular descriptors of the local atomic environment) onto an energy value. Following Marchand \textit{et al.}~\cite{marchand2020machine}, the potential uses a $64$--$24$--$24$--$1$ architecture with two hidden layers of 24 neurons each, 
\begin{equation}
E_i^{\mathrm{NNP}} = 
f_3\!\left[
b_3 + \sum_{k=1}^{24} a^{(2,3)}_{k,1}\,
f_2\!\left(
b^{(2)}_k + \sum_{j=1}^{24} a^{(1,2)}_{j,k}\,
f_1\!\left(
b^{(1)}_j + \sum_{m=1}^{64} a^{(0,1)}_{m,j} G_m
\right)\right)\right].
\end{equation}
Here $a^{(q,p)}_{j,k}$ is the weight connecting neuron $j$ in layer $q$ to neuron $k$ in layer $p$, and $b_j^{(q)}$ is the bias of neuron $j$ in layer $q$. The activation functions are chosen as
$f_1(x) = \ln(1+e^x)$ (softplus) for the first hidden layer, and $f_2(x)=f_3(x)=x$ (identity) for the second hidden layer and output layer, so that the only nonlinearity enters through $f_1$. The input features $G_m$ are Behler--Parrinello symmetry functions, combining radial and angular terms to encode the local environment in a rotationally and permutationally invariant way. The specific parameter sets and definitions follow Refs.~\cite{behler2007generalized,marchand2020machine}.

The NNP is trained by minimizing a combined loss function over DFT energies and forces for every configuration in the training database, as implemented in \texttt{n2p2}~\cite{singraber2019parallel}. The loss function sums the squared errors in energies and atomic force components, with a weighting factor applied to the force term to balance its influence relative to the energies. Forty realizations of the NNP were fit to a comprehensive DFT dataset of $\sim$4857 Al--Cu configurations, including pure elements, ordered intermetallics, point defects, surfaces, stacking faults, and coherent precipitate--matrix interfaces~\cite{marchand2020machine}. Among these, the NNP11  parametrization was selected because it yielded the lowest error in reproducing the aluminum shear modulus $C_{44}$, a key quantity for capturing dislocation glide and shear-driven interactions with GP-zones. Within its training and testing dataset, NNP11 is reported to achieve near-DFT accuracy~\cite{marchand2020machine}.

\subsubsection*{Shear Loading}

Prior to mechanical loading, the energy of all configurations was minimized. Quasi-static shear loading was then applied by increasing the target shear stress in increments of \SI{5}{\mega\pascal}. For each increment, the prescribed shear stress \(\tau_{xy}\) was converted into a per-atom force that was applied to atoms in the top and bottom regions (with respect to Y) of the simulation cell,
\begin{equation}\label{eq-forcetop}
f_{\text{top}} = \frac{\tau_{xy}A_{xz}}{N}, \qquad
f_{\text{bottom}} = -\frac{\tau_{xy}A_{xz}}{N},
\end{equation}
where \(A_{xz}\) is the cross-sectional area normal to the loading direction and \(N_{\mathrm{top}}\) and \(N_{\mathrm{bottom}}\) are the number of atoms in the regions over which the traction is applied. Each loading increase was followed by energy minimization using the nonlinear conjugate-gradient (\texttt{cg}) algorithm with a force tolerances of \(10^{-4}\)\,eV\,\AA$^{-1}$. The athermal critical resolved shear stress (CRSS) was taken as the value of the applied load at which the dislocation successfully overcame the GP-zone.

To assess the directional dependence of the dislocation–precipitate interaction, two loading directions were considered, corresponding to opposite signs of the applied shear stress $\tau_{xy}$. The sign of the applied shear determines the direction from which the dislocation approaches the GP-zone and consequently, the side of the dislocation,tensile (below the slip plane) or compressive (above the slip plane), that first interacts with the GP-zone. 

For the GP $0^\circ$ orientation, the platelet is symmetric with respect to the slip plane; therefore, the strengthening is symmetric with respect to the sign of $\tau_{xy}$, i.e. the direction from which the dislocaiton approaches the precipitate. In contrast, for the GP $60^\circ$ orientation, the interaction is asymmetric. For the configuration we have simulated, the compressive lobe of the edge dislocation interacts first with the GP-zone upon the application of a positive $\tau_{xy}$ shear; and correspondingly, the tensile lobe of the dislocation interacts first with the GP zone upon the application of a negative $\tau_{xy}$ shear. We will refer to these cases as forward and backward shear, respectively. 

\subsubsection*{Cu$\leftrightarrow$Al exchanges}

Single Cu$\leftrightarrow$Al exchanges were examined under two loading regimes: at zero applied shear stress and a stress value below the CRSS where dislocation motion is being obstructed by the GP-zone. Each possible exchange is defined between a matrix site $i$ and a GP-zone site $j$, so that a forward exchange $i\!\to\!j$ and its reverse $j\!\to\!i$ form a paired event. For each exchange, a subsequent force minimization is performed using the nonlinear conjugate-gradient algorithm (tolerance \(10^{-4}\,\mathrm{eV\,\text{\AA}^{-1}}\)). The minimization provides a post-exchange energy $W_1$ that can be compared to the pre-exchange energy, $W_0$, to obtain a thermodynamic driving force for the exchange 
\[
\Delta W_{ij{}} = W_1 - W_0.
\]
A negative value, $\Delta W_{ij}<0$, indicates that the exchanged configuration is energetically favored; a positive value, $\Delta W_{ij}>0$, indicates that it is unfavorable.

All nearest-neighbor Cu$\leftrightarrow$Al exchanges involving Al atoms within one atomic layer of the GP-zone were systematically sampled, providing a complete mapping of local Cu atom-exchange energetics.

For the energetically favorable exchanges with $\Delta W_{ij} < 0$, quasi-static shearing was resumed from the swapped configuration to quantify the extent to which the exchange altered the critical resolved shear stress (CRSS). 

\subsubsection*{Visualization and Post-Processing}

All structural analyses and visualizations were performed using the OVITO software package~\cite{stukowski2009visualization}. Atoms were colored by their potential energy or atomic species. The Dislocation Extraction Algorithm (DXA) was tested, but it proved unreliable in configurations where partial dislocations where near the GP zones; consequently, dislocation lines were traced manually from the atomic shear strains (relative to the initial configuration). This approach provided clear visual evidence of dislocation glide, GP-zone interactions, and the Cu$\leftrightarrow$Al exchange events. 

\subsubsection{Kinetic Monte Carlo}

The kinetic Monte Carlo (kMC) framework transforms the catalog of single-hop Cu$\leftrightarrow$Al exchanges characterized by energy change $\Delta W_{ij}$ and change in CRSS $\Delta \tau_{ij}$ extracted from the atomistic simulations into a time-resolved trajectory of Cu atom rearrangement during the dislocation overcoming the GP zone. 

Each event $m$ corresponds to a single exchange, which may occur in the forward direction $i \to j$, producing an energy change $\Delta W_{ij}$ and a corresponding change in CRSS $\Delta\tau_{ij}$, or in the reverse direction $j \to i$, which yields changes of equal magnitude and opposite sign.

\subsubsection*{Activation Barriers and Transition Rates}
To convert the $\Delta W_{ij}$ values derived from atomistic simulations into physical transition rates, we combine them with a reference migration barrier $\Delta E_0$ that sets the activation barrier for exchange. The forward and backward activation barriers are constructed using the symmetric partition 
\begin{equation}
  E_{f,ij} = \Delta E_0 + \tfrac{1}{2}\Delta W_{ij}, \qquad
  E_{b,ij} = \Delta E_0 - \tfrac{1}{2}\Delta W_{ij},
\end{equation}
which guarantees detailed balance. In this formulation, $\Delta W_{ij}$ is a thermodynamic bias between the two configurations, while $\Delta E_0$ is the activation enthalpy for a hop in the absence of any thermodynamic preference. This construction follows the model of Curtin \textit{et al.}~\cite{curtin2006predictive}, where the reference barrier $\Delta E_0$ corresponds to the single hop migration barrier in the dislocation core, and the thermodynamic energy difference $\Delta W_{ij}$ shifts the forward and backward barriers by $\pm\Delta W_{ij}/2$. 

The rate for exchanges is modeled with Arrhenius kinetics,
\begin{equation}
  k_{f,ij} = \nu_0 \exp\!\left[-\frac{E_{f,ij}}{k_\mathrm{B}T}\right],
  \qquad
  k_{b,ij} = \nu_0 \exp\!\left[-\frac{E_{b,ij}}{k_\mathrm{B}T}\right],
\end{equation}
where $\nu_0$ is the attempt frequency and $k_{\mathrm{B}}$ is Boltzmann’s constant (eV/K). 

\subsubsection*{Event Exclusivity}

Each GP site can accommodate only a single incoming atom. Therefore, the kMC state is defined by the occupancy of the GP zone sites. A site is either free or occupied by exactly one event $m$, which determines whether forward, backward, or no transitions to that site are permitted.

\subsubsection*{Time Evolution: Gillespie Algorithm}

The single-hop Cu$\leftrightarrow$Al exchanges are simulated using a kMC scheme based on the Gillespie direct stochastic simulation algorithm ~\cite{gillespie1976general,gillespie1977exact}, in which the system evolves as a continuous-time Markov process with event rates derived from atomistic simulations' energy barriers.

At each step, two random numbers, $u_1$ and $u_2$, are drawn from a uniform distribution (0,1). Then a random waiting time is calculated, $\Delta t = -\frac{\ln u_1}{R(t)}$ where $R=\sum_n k_n$ is the sum of transition rates for each possible exchange. The particular exchange event is selected proportional to its rate by finding the smallest index $n^*$ satisfying $\sum_{i=1}^{n^*} k_i \ge u_2 R$. Thus, $u_1$ governs when the next transition occurs, while $u_2$ determines which event is executed. Executing event $n^*$ updates both the GP-site occupancy and the cumulative strengthening depending on direction. The possible event set is then updated and the process repeated advancing the simulation time. The process is continued until a prescribed final time, chosen such that the strengthening response converges to a saturated value corresponding to the steady-state configuration of alloying atoms redistribution.

\subsubsection{Dynamic Strain Aging Model}\label{sec:dsa}
We employ the Soare–Curtin single–mechanism framework for dynamic strain aging (DSA), where plastic flow is governed by the thermally activated escape of a pinned dislocation over a time- and stress-dependent energy barrier~\cite{soare2008single}. The plastic strain rate $\dot{\varepsilon}$ follows the Orowan relation,
\begin{equation}
\dot{\varepsilon}(\tau)=\frac{\Omega}{t_w(\tau)},
\label{eq:orowan}
\end{equation}
where the mean waiting time is given by
\begin{equation}
t_w(\tau)=\int_{0}^{\infty}\exp\Big[-\int_{0}^{t} v(\tau,t'),dt'\Big],dt,
\label{eq:tw}
\end{equation}
and the activation rate is
\begin{equation}
v(\tau,t)=\nu_0 \exp\left[-\frac{\Delta F(\tau,t)}{kT}\right].
\label{eq:rate}
\end{equation}
The activation barrier is expressed as
\begin{equation}
\Delta F(\tau,t)=\Delta F_0\left(1-\frac{\tau-\Delta\tau_a(t)}{\tau_0}\right)^{\alpha},
\label{eq:barrier}
\end{equation}
with the time dependence taking the form of an aging-induced back-stress  described by
\begin{equation}
\Delta\tau_a(t)=\Delta\tau_{\infty}\left[1-e^{-(t/t_{\mathrm{sat}})^{n}}\right].
\label{eq:aging}
\end{equation}

Here, $\Omega$ is the increment in strain associated with the escape of a pinned dislocation, $\nu_0$ is the attempt frequency, $\Delta F_0$ is the zero-stress barrier for dislocation escape from the GP zone, $\alpha$ is the barrier-shape exponent, $\tau_0$ the athermal strength (CRSS at 0 K), and $\Delta\tau_a(t)$ the aging back-stress determined by a diffusion time constant $t_{\mathrm{sat}}$, a saturation strength $\Delta\tau_{\infty}$, and a kinetic exponent $n$. 

The mean waiting time $t_w(\tau)$ is evaluated numerically, and the strain-rate sensitivity is obtained as
\begin{equation}
m(\dot{\varepsilon})=\frac{d\ln \tau}{d\ln \dot{\varepsilon}},
\label{eq:m_general}
\end{equation}
\subsection{Parameter Selection}
\subsubsection{Parameters for Kinetic Monte Carlo}
To apply the kMC model to the present Al--Cu system, a representative activation barrier for cross-core solute migration must be specified. Atomistic reference data for such processes are most comprehensively available for Mg in fcc Al. Picu \textit{et al.}~\cite{picu2004atomistic} computed activation enthalpies for Mg migration along dislocation cores. Their lowest orientation-dependent core value, $\Delta H_c \approx 0.82$~eV for the $60^\circ$ dislocation, reflects the energetic cost of alloy atom exchanges within the dislocation core. Several observations support the choice of a comparable activation barrier for exchange in Al-Cu. First, the minimum migration barriers in the dislocation core are generally thought to be $0.5$--$0.6$ of bulk values~\cite{balluffi1970measurements,zhu1991activation}, and DFT calculations by Simonovic \textit{et al.}~\cite{simonovic2009impurity} report a bulk migration barrier of $\sim$1.18~eV for Cu in Al. Second, barrier for Cu in Al would be expected to be similar given the similarity of misfit volume. Taken together, these considerations make an effective core migration barrier in the range $0.7$--$0.9$~eV reasonable and we therefore adopt $\Delta E_0 = 0.8$~eV in the kMC simulations.
 
The kMC simulation outputs a time dependent change in strengthening $\Delta\tau(t)$. A simple two-parameter sigmoidal relation aging law used in ~\cite{curtin2006predictive} (Eq.~\eqref{eq:aging}) has been fitted to the kMC data to capture the gradual transition from an initial transient regime to a quasi-steady plateau. The time at which the curve enters this plateau defines the characteristic aging time $t_{d}=t_{\mathrm{sat}}$, while the asymptotic value of the curve corresponds to the steady-state strengthening $\Delta\tau_{\infty}$.

\subsubsection{Parameters for DSA Model}

To calibrate the DSA model for the Al--Cu alloy, the quantities $\Delta\tau_{\infty}/\tau_0$ and $t_d$ are extracted directly from the kMC simulations. The remaining model parameters, the attempt frequency $\nu_0$, the zero-stress activation barrier $\Delta F_0$, and the barrier-shape exponent $\alpha$, are chosen following Saroukhani \textit{et al.}~\cite{saroukhani2016harnessing}, i.e. $\nu_0 = 1.54\times10^{13}$~s$^{-1}$ and $\Delta F_0 = 1.7$~eV for a dislocation overcoming a weak obstacle in periodic atomistic simulations. The barrier-shape exponent is taken as $\alpha = 3/2$, a standard choice for smoothly varying energy landscapes~\cite{kocks1975thermodynamics,olmsted2006molecular,curtin2006predictive}.

The remaining input to the DSA model is the Orowan strain increment,
\begin{equation}
  \Omega = \rho_m\, b\, \rho_f^{-1/2},
\label{eq:omega}
\end{equation}
where $\rho_m$ is the mobile dislocation density, $b$ is the Burgers vector, and $\rho_f^{-1/2}$ is the mean spacing between obstacles (precipitates or forest dislocations). 
Assuming precipitate-limited glide with obstacle spacing $L_p = \rho_f^{-1/2}$, we adopt the values reported by Singh\textit{ et al.}~\cite{singh2013atomistic}: $\rho_m = 10^{10}$~m$^{-2}$, $L_p = 15.9$~nm ($1.59\times10^{-8}$~m), and $b = 2.86$~\AA ($2.86\times10^{-10}$~m). These values yield an Orowan increment of $\Omega = 4.55\times10^{-8}$. 

\section{Results and Discussion}

\subsection{Stability of Isolated GP Zone}\label{sec:farfield}

\begin{figure}[!h]
\centering
\includegraphics[width=1\textwidth]{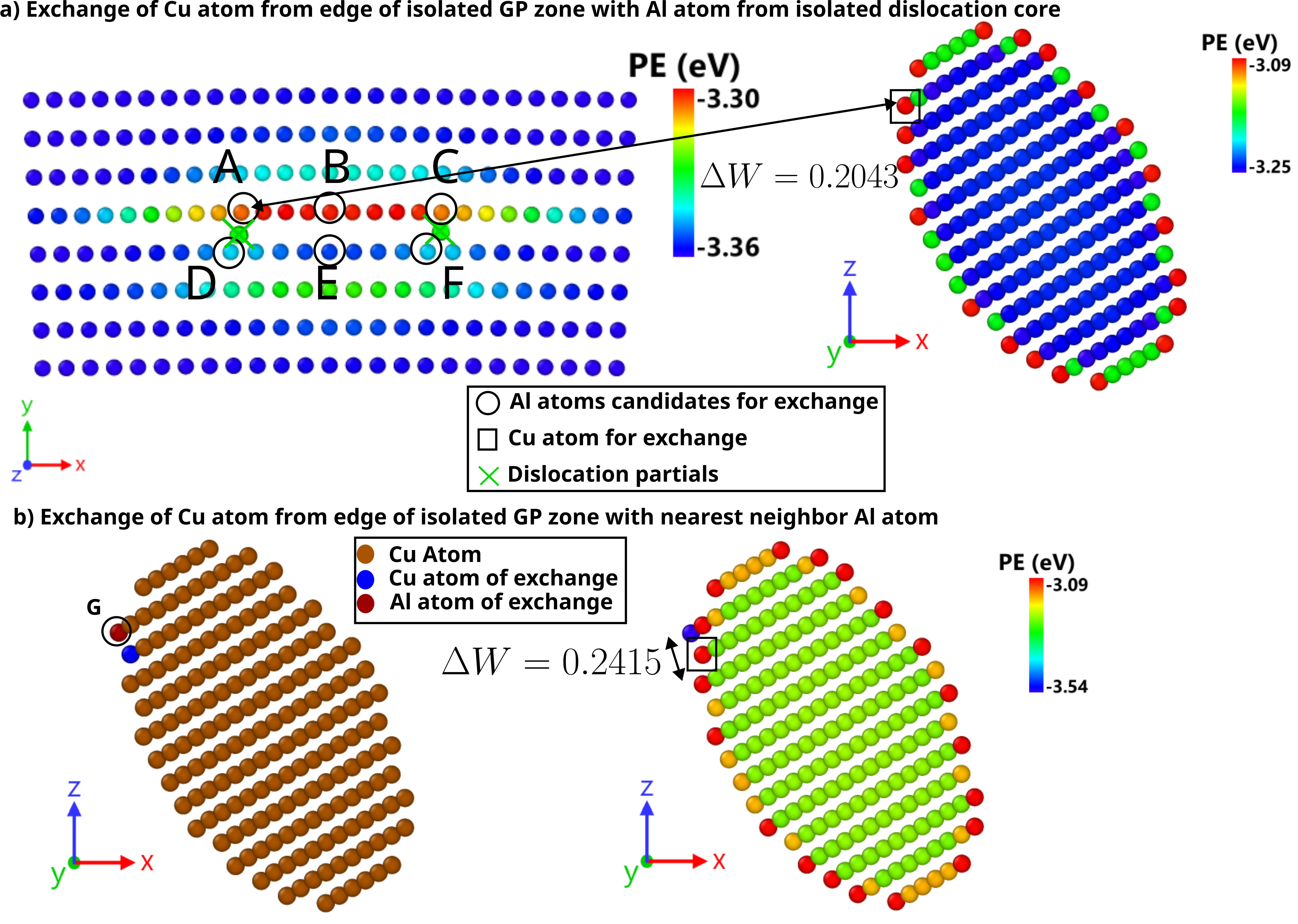}
\caption{Cu$\leftrightarrow$Al exchanges when the dislocation is far from the GP zone in the Al--Cu alloy. (a)~Exchange between a Cu atom from the GP zone and Al atoms at the dislocation core. Six candidate Al sites (A--F) are selected based on their positions relative to the compressive and tensile sides of the dislocation core. The Cu atom chosen for the exchange corresponds to the site with the highest potential energy within the GP zone, resulting in the Cu atom occupying the GP-zone boundary, and the configuration is selected to minimize the energy difference $\Delta W_{ij}$. (b)~Single-hop (nearest-neighbour) Cu$\leftrightarrow$Al exchange within the GP zone and its surroundings. The Al atom~G~is selected to minimize~$\Delta W_{ij}$. The corresponding $\Delta W_{ij}$ values for all labeled sites (A--G) are summarized in Table~\ref{tab:deltaWvalues}.}
\label{fig0}
\end{figure}

To establish a baseline, we evaluated a series of Cu$\leftrightarrow$Al exchanges involving Al atoms located either along the core of an edge dislocation positioned sufficiently far from the GP zone (to eliminate any GP zone-dislocation interaction (Fig.~\ref{fig0}a)), or within the Al matrix immediately adjacent to the GP zone accessible through a single-hop exchange (Fig.~\ref{fig0}b). For each GP zone orientation, the Cu atom selected for exchange is the site with the highest potential energy within the GP zone, which consequently yields the minimum exchange energy difference, $\Delta W_{ij,\text{min}}$. 

To sample the range of local environments generated by the dissociated dislocation core, six Al sites (A–F) were selected: three on the tensile side (D-F) and three on the compressive side (A-C), spanning the leading partial, the trailing partial, and the region between them. An additional site (G) was chosen at the GP-zone interface. Site G corresponds to the Cu$\leftrightarrow$Al exchange that has the lowest value of $\Delta W_{ij}$ at the GP zone interface. 

\begin{table}[h!]
\centering
\caption{Energy differences $\Delta W_{ij}$ (in eV) associated with Cu$\leftrightarrow$Al exchanges at locations near the dislocation core and GP zone, obtained using the ADP and NNP potentials. The locations A--G correspond to the Al atomic sites shown in Fig.~\ref{fig0}. 
All exchanges are energetically unfavorable $\Delta W_{ij}>0$. }
\label{tab:deltaWvalues}
\begin{tabular}{lcccc}
\hline
\textbf{Location} & \textbf{GP (ADP)} & \textbf{GP (NNP)} \\
\hline
A & 0.204 & 0.054 \\
B & 0.222 & 0.163 \\
C & 0.208 & 0.055 \\
D & 0.398 & 0.305 \\
E & 0.663 & 0.313 \\
F & 0.296 & 0.313 \\
G & 0.242 & 0.094 \\
\hline
\end{tabular}
\end{table}

Table~\ref{tab:deltaWvalues} shows the energy differences $\Delta W_{ij}$ associated with each of these Cu$\leftrightarrow$Al exchanges. In every case, the energy change is positive, indicating that exchanging a Cu atom in the GP zone with an Al atom, either at the distant dislocation core or in the local matrix, incurs an energy penalty. The ADP potential predicts $\Delta W_{ij}$ values in the range $0.20$--$0.66$~eV, while the NNP potential yields lower values of $0.05$--$0.31$~eV. As might be expected from the mismatch in atomic volume, exchanges placing Cu atoms on the compressive side of the dislocation core are more favorable than those on the tensile side of the core. These more favorable exchanges involving the compressive side have similar energetics to the nearest neighbor exchange G at the GP-zone interface. 

The values reported in Table~\ref{tab:deltaWvalues} depend on the precise edge configuration of the GP zone and the manner in which it is constructed. In study of a second configuration not reported here, the results varied by  0.05~eV, a level small enough not to qualitatively affect our findings. 

\newpage
\subsection{Nearest-Neighbour Exchanges of Cu with Al when Dislocation is Pinned at GP Zone}

When the dislocation is pinned at the GP zone, its dissociated core generates a heterogeneous local strain field in the immediate vicinity of the GP zone. This local environment can alter the relative stability of Cu and Al atomic sites near the GP zone. To quantify how the presence of a pinned dislocation modifies the thermodynamic driving forces for alloy atom rearrangement, we performed a systematic set of single-hop Cu$\leftrightarrow$Al exchanges for every Cu atom within the GP zone and for all of its nearest-neighbour Al atoms within a cutoff of \SI{3.2}{\angstrom}. All simulations were performed at applied stress levels such that the dislocation is pinned at the GP zone and has not yet overcome it. We therefore begin by determining the CRSS of the GP zone in the absence of alloy atom exchanges.

\subsubsection{Athermal Critical Resolved Shear Stress} \label{sec:crss}

The athermal critical resolved shear stress (CRSS) required for an edge dislocation to overcome a GP zone is first determined to identify appropriate loading shear stresses for the Cu$\leftrightarrow$Al exchange simulations and to establish a reference strength against which the effect of alloy atom rearrangements can be compared. To obtain these athermal CRSS values, we performed 0 K molecular-statics loading simulations for both GP~0$^\circ$ and GP~60$^\circ$ orientations using the ADP and NNP11 interatomic potentials. The resulting CRSS are summarized in Table~\ref{tab:gpstrengths}, alongside the atomistic values reported by Singh and Warner~\cite{singh2010mechanisms}.

\begin{table}[h!]
\centering
\caption{Athermal CRSS (in MPa) for edge-dislocation overcoming the GP zones at 0~K using ADP and NNP11 potentials, compared with atomistic simulations values reported by Singh and Warner ~\cite{singh2010mechanisms}. The angle and tensile/compressive nomenclature corresponds to the orientation and direction of loading as described in the Methods section.}
\label{tab:gpstrengths}
\begin{tabular}{lccc}
\hline
\textbf{Case} & \textbf{Singh et al.~\cite{singh2010mechanisms}} & \textbf{our simulations} \\
\hline
GP 0$^\circ$, ADP & 185 & 175\\
GP 60$^\circ$ (tensile), ADP & 232 & 235\\
GP 60$^\circ$ (compressive), ADP & -- & 255\\
\hline
GP 0$^\circ$, NNP11 & -- & 60 \\
GP 60$^\circ$ (tensile), NNP11 & -- & 70\\
GP 60$^\circ$ (compressive), NNP11 & -- & 125\\
\hline
\end{tabular}
\end{table}

For the ADP potential, the athermal response is first examined for the GP 0$^\circ$ orientation. The simulations give $\tau_{\mathrm{ath}}^{0^\circ} \approx 175~\mathrm{MPa}$,
in close agreement with the \(185~\mathrm{MPa}\) reported by Singh and Warner~\cite{singh2010mechanisms}. After the GP zone has been overcome by the dislocation, the resulting configuration involves a dislocation loop, yet the bow-out angle prior to this was much smaller than the \(90^\circ\) expected for a classical Orowan looping process. This behaviour matches the loop-like states reported in earlier atomistic studies. Esteban-Manzanares \textit{et al.}~\cite{esteban2019strengthening} showed that such behavior arises when doing athermal energy minimization and may correspond to a shallow energy minimum, suggesting that they are not be indicative of real-world behavior.

For the GP~60$^\circ$ orientation with the ADP potential, the CRSS depends on whether the GP zone first interacts with the tensile or compressive lobe of the dislocation stress field as described in the methods section. In the tensile case, the dislocation overcomes the GP-zone by cutting it with an athermal CRSS of $\tau_{\mathrm{ath}}^{60^\circ} \approx 235~\mathrm{MPa}$, consistent with the \(232~\mathrm{MPa}\) cutting strength reported by Singh and Warner~\cite{singh2010mechanisms}. When the sign of the applied stress is reversed, the GP zone is first acted upon by the compressive lobe of the dislocation. In this case the GP zone acts as a slightly stronger obstacle, requiring \(255~\mathrm{MPa}\) for the dislocation to overcome it. In this configuration, the dislocation forms a loop around the platelet, yet the loop is shallow and does not exhibit the large bow-out angles characteristic of a full Orowan process. This behaviour resembles what was observed for the GP\,0$^\circ$ orientation, although in the present case the dislocation bows out more steeply, indicating that the GP zone is acting as a stronger obstacle. 

Relative to the above ADP potential results, simulations with the NNP11 potential predict substantially lower athermal strengths and different mechanisms in most cases, except for the GP~60$^\circ$ tensile orientation. For GP~60$^\circ$, both the tensile and compressive configurations lead to cutting, with corresponding strengths of approximately \(70~\mathrm{MPa}\) and \(125~\mathrm{MPa}\), respectively. For the GP~0$^\circ$ zone orientation, a cutting process occurs with an athermal strength of $\tau_{\mathrm{ath}}^{0^\circ,\mathrm{NNP}} \approx 60~\mathrm{MPa}$.

\subsubsection{Cu$\leftrightarrow$Al Nearest-Neighbour Exchange Energies}

\begin{figure}[!h]
\centering
\includegraphics[width=\textwidth]{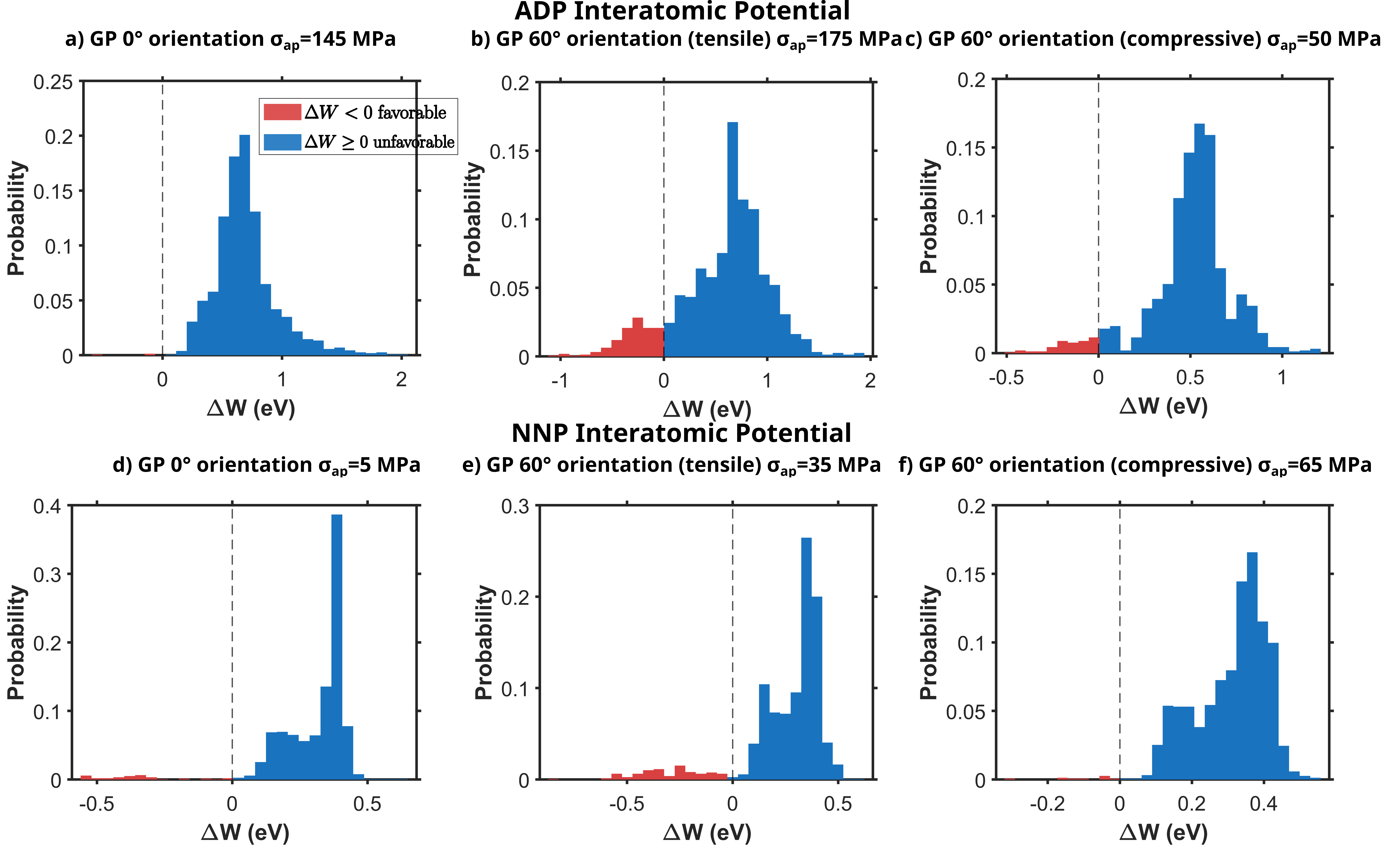}
\caption{Probability distributions of energy changes $\Delta W_{ij}$ for all Cu$\leftrightarrow$Al nearest-neighbour exchanges when the dislocation is pinned at the GP zone. Blue indicates unfavourable exchanges ($\Delta W_{ij}\ge0$), while red highlights favourable exchanges ($\Delta W<0$). The applied shear stress $\sigma_{\mathrm{ap}}$ for each configuration is indicated in the panel titles.}
\label{fig:hist}
\end{figure}

Using both the ADP and NNP11 interatomic potentials, the exchange energies $\Delta W_{ij}$ were computed for all Cu$\leftrightarrow$Al nearest-neighbour exchanges when the dislocation is pinned at the GP zone at a stress below the CRSS. Histograms of the exchange energies are shown in Fig.~\ref{fig:hist}. In all cases, the tail of the $\Delta W_{ij}$ distributions contains negative-$\Delta W_{ij}$ values that were not present for Cu$\leftrightarrow$Al exchanges involving a distant dislocation (discussed in section~\ref{sec:farfield}). These negative values correspond to thermodynamically favorable Cu$\leftrightarrow$Al exchanges that appear in the presence of a pinned edge dislocation at the GP zone. 

When using the ADP potential for the GP\,0$^\circ$ orientation under moderate shear ($\sigma_{\mathrm{ap}} = 145~\text{MPa}$), only a very small fraction of exchange events ($\approx 0.19\%$) are energetically favorable ($\Delta W_{ij} < 0$) (Fig.~\ref{fig:hist}a), The most negative exchange energy was $\Delta W_{ij,\min} = -0.6~\text{eV}$. In this configuration, the energetically favorable Cu$\leftrightarrow$Al exchanges are aligned with the dislocation line direction, rendering them independent of the dislocation stress field. 

With the NNP11 potential, the fraction of favorable exchanges increases to $3.31\%$, with $\Delta W_{ij,\min} = -0.56~\text{eV}$ and a mean value of $\Delta W_{ij,\mathrm{mean}} = -0.36~\text{eV}$. The substantial difference between the NNP11 and ADP $\Delta W_{ij}$ exchange energy distributions for this orientation is not surprising in light of the difference in the dislocation bypass mechanisms between the potentials, i.e. cutting vs looping. This result is consistent with the simulations of Wu \textit{et al.}, who showed that Cu stabilization is favored with cutting but not looping mechanisms. Furthermore, Chen \textit{et al.} investigated the influence of cutting versus non-cutting precipitates on the Portevin--Le Chatelier effect in Al--Mg--Sc--Zr alloys. Their results demonstrates that the precipitation strengthening mechanism, cutting or looping, strongly governs the local availability of solutes and vacancies near dislocations~\cite{chen2021influence}. In particular, precipitate cutting was observed to release alloying atoms and vacancies, whereas interactions with non-cutting precipitates did not exhibit this behavior. 

For the GP\,60$^\circ$ orientation, the exchange energetics are strongly dependent on the direction of shear. With the ADP potential and the GP\,60$^\circ$ tensile configuration, at $\sigma_{\mathrm{ap}} = 175$~MPa the leading partial has sheared through the GP zone and the trailing partial remains pinned. In this case, approximately 12\% of nearest-neighbor exchanges are energetically favorable (Fig.~\ref{fig:hist}b), with $\Delta W_{ij,\min} = -0.512~\text{eV}$ and $\Delta W_{ij,\mathrm{mean}} = -0.157~\text{eV}$. In contrast, for the ADP GP\,60$^\circ$ compressive configuration, where neither dislocation partials has overcome the precipitate, the fraction of favorable exchanges is smaller, on the order of 5\% (Fig.~\ref{fig:hist}c), with $\Delta W_{ij,\min} = -1.119~\text{eV}$ and $\Delta W_{ij,\mathrm{mean}} = -0.288~\text{eV}$. 

The NNP11 potential yields substantially narrower $\Delta W_{ij}$ distributions and far fewer large-magnitude negative exchanges. For the tensile GP\,60$^\circ$ configuration, approximately 8.6\% of exchanges are favorable, with $\Delta W_{ij,\min} = -0.873~\text{eV}$ and $\Delta W_{ij,\mathrm{mean}} = -0.295~\text{eV}$. For the GP\,60$^\circ$ compressive configuration, only about 0.65\% of exchanges are favorable, with $\Delta W_{ij,\min} = -0.320~\text{eV}$ and $\Delta W_{ij,\mathrm{mean}} = -0.103~\text{eV}$. Importantly, both potentials predict the same qualitative orientation dependence, with the GP\,60$^\circ$ zone being more susceptible to dislocation-assisted Cu$\leftrightarrow$Al nearest-neighbour exchanges than the GP\,0$^\circ$ zone. For the \,60$^\circ$ configuration, such favorable exchanges occur for both tensile and compressive configurations with the ADP potential (Fig.~\ref{fig:hist}b,c).

The majority of favourable exchanges fall within $0.1$–$0.4$ eV. For context, we note that these values are much smaller than typical pipe-diffusion activation energies reported for Al-based alloys $0.8$–$0.9$ eV~\cite{picu2004atomistic} and the effective bulk diffusion activation energy for Cu in Al ($1.2$ eV) as estimated from DFT vacancy-formation and migration data~\cite{simonovic2009impurity}. The most favourable exchanges predicted occur almost exclusively in the GP 60$^\circ$ cutting geometry, where the intersecting dislocation core produces significant local shear and pronounced tension–compression asymmetry. Overall, these results demonstrate that a dislocation pinned at a GP zone can enhance the local redistribution of alloy atoms, consistent with previous experimental observations of dislocation-assisted transport~\cite{suzuki1962segregation,legros2005pipe}. 

\subsubsection{Spatial Maps of Favourable Exchanges within and around the GP Zone}

\begin{figure}[!h]
\centering
\includegraphics[width=0.9\textwidth]{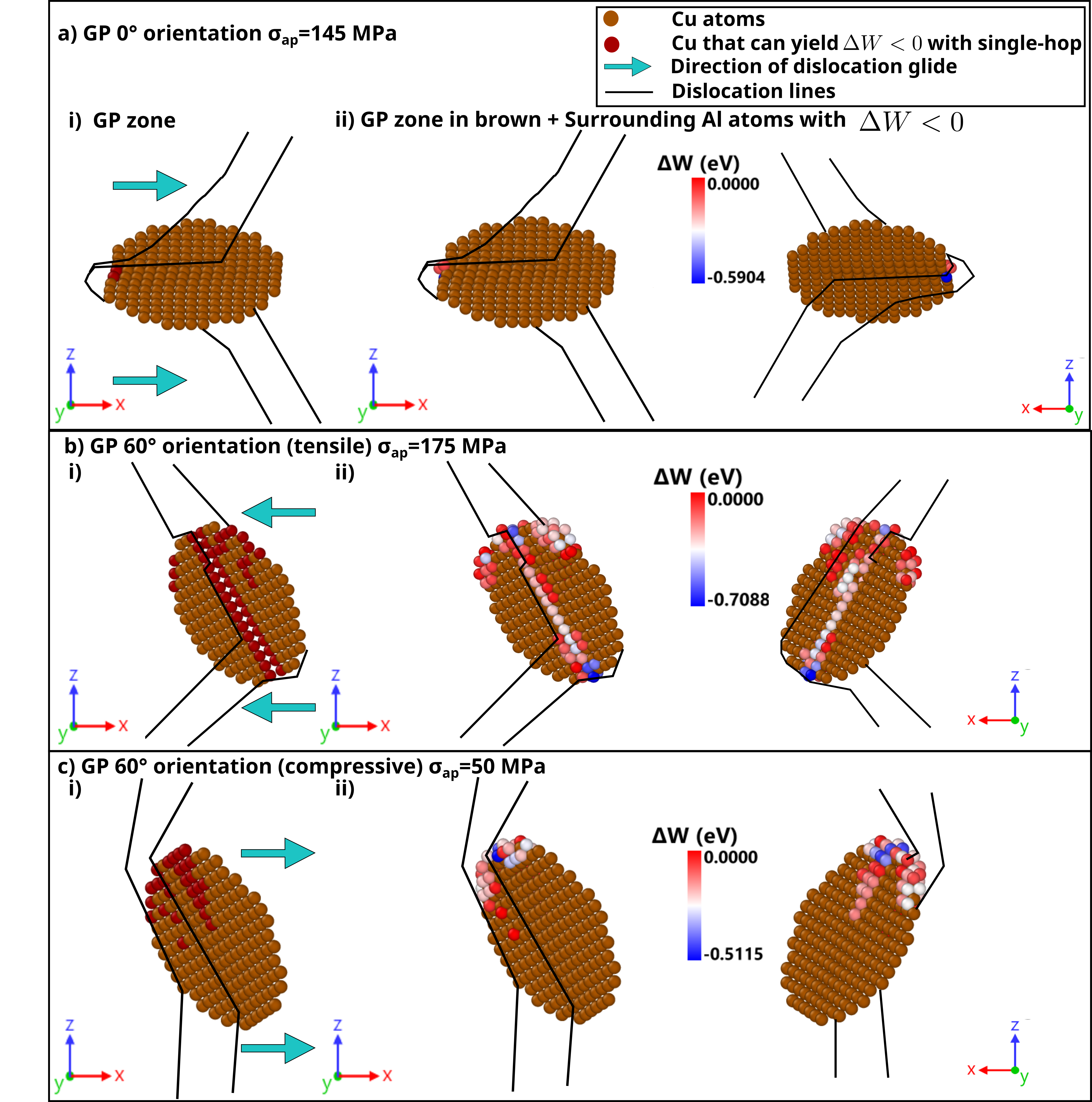}
\caption{Energetically favourable Cu$\leftrightarrow$Al nearest-neighbour exchanges for the ADP potential. Panel (i) shows Cu atoms coloured according to if they have any energetically favorable single-hop exchanges, $W_{ij,\mathrm{min}}<0$. Panel (ii) shows Al atoms from two perspectives that are associated with favorable exchanges and colored according to their most favorable exchange energy. }
\label{fig2}
\end{figure}

Fig.~\ref{fig2} shows the spatial distribution of favourable Cu$\leftrightarrow$Al nearest-neighbour exchanges for the ADP potential. Cu atoms inside the GP zone are coloured according to if they have any energetically favorable single-hop exchanges $\Delta W_{ij,\mathrm{min}}<0$. Al atoms associated with favorable exchanges are also shown and colored according to their lowest exchange energies. 

For the ADP potential and the GP 0$^\circ$ configuration (Fig.~\ref{fig2}a), favourable exchanges only occur in a localized region at the edge of the GP zone that is near to the bowed dislocation. 

For the GP 60$^\circ$ orientation, the result depends on the direction of dislocation glide (Fig.~\ref{fig2}b,c). For the 60$^\circ$ tensile configuration, the tensile field of the gliding dislocation first interacts with the precipitate, which is then overcome via a cutting mechanism. Exchanges were examined at the state where the leading Shockley partial has sheared through the lower half of the GP zone and the trailing partial remains anchored at the edge of the GP zone. Favourable Cu sites are positioned continuously along the line where the leading partial has sheared the GP zone and created a step. The corresponding favorable Cu$\leftrightarrow$Al exchange sites are distributed nearly symmetrically across the slip plane. The most negative $\Delta W_{ij}$ values are located near the GP zone edges on the slip plane. 

For the ADP potential and the GP 60$^\circ$ compressive configuration (Fig.~\ref{fig2}c) under forward shear, neither partial dislocation has cut the GP zone in the examined configuration. Accordingly, the favorable exchanges are localized near the location where the partial dislocations impinge on the edge of the GP zone. 

The results presented above suggest that dislocation-assisted Cu$\leftrightarrow$Al exchange events in Al--Cu alloys are localized and strongly dependent on precipitate orientation and dislocation interaction mode. Across both ADP and NNP11 potentials, energetically favorable exchanges occur in the immediate vicinity of the dissociated dislocation core, most notably near the leading partial in the GP $60^\circ$ configurations. This localization indicates that the dominant driver of alloy atom rearrangement is the strongly perturbed atomic environment associated with the dislocation core and its stress field.

The localization of favorable exchange events naturally invites comparison with prior studies of solute transport and stabilization at crystalline defects. Continuum modeling by Epperly and Sills~\cite{epperly2020comparison} has shown that cross-core diffusion can arise near extended dislocations due to locally reduced activation barriers within the dissociation region. While the present simulations do not explicitly resolve diffusion kinetics, the confinement of negative-$\Delta W_{ij}$ events to the dislocation core region is consistent with this picture of core-mediated solute mobility. Dislocation cores are also known to act as fast transport pathways for solutes, a phenomenon commonly referred to as dislocation pipe diffusion. Such behavior has been reported extensively in both experimental measurements and atomistic simulations across a wide range of alloy systems~\cite{volin1971measurement,picu2004atomistic,legros2005pipe,legros2008observation}. Building on these concepts, recent atomistic predictions by Turlo and Rupert~\cite{turlo2020prediction} have further demonstrated that sustained solute enrichment at dislocation cores can stabilize distinct chemical–structural defect states, often termed linear complexions.

\subsubsection{Strength Change due to Cu$\leftrightarrow$Al Nearest-Neighbour Exchanges}

\begin{figure}[!h]
\centering
\includegraphics[width=\textwidth]{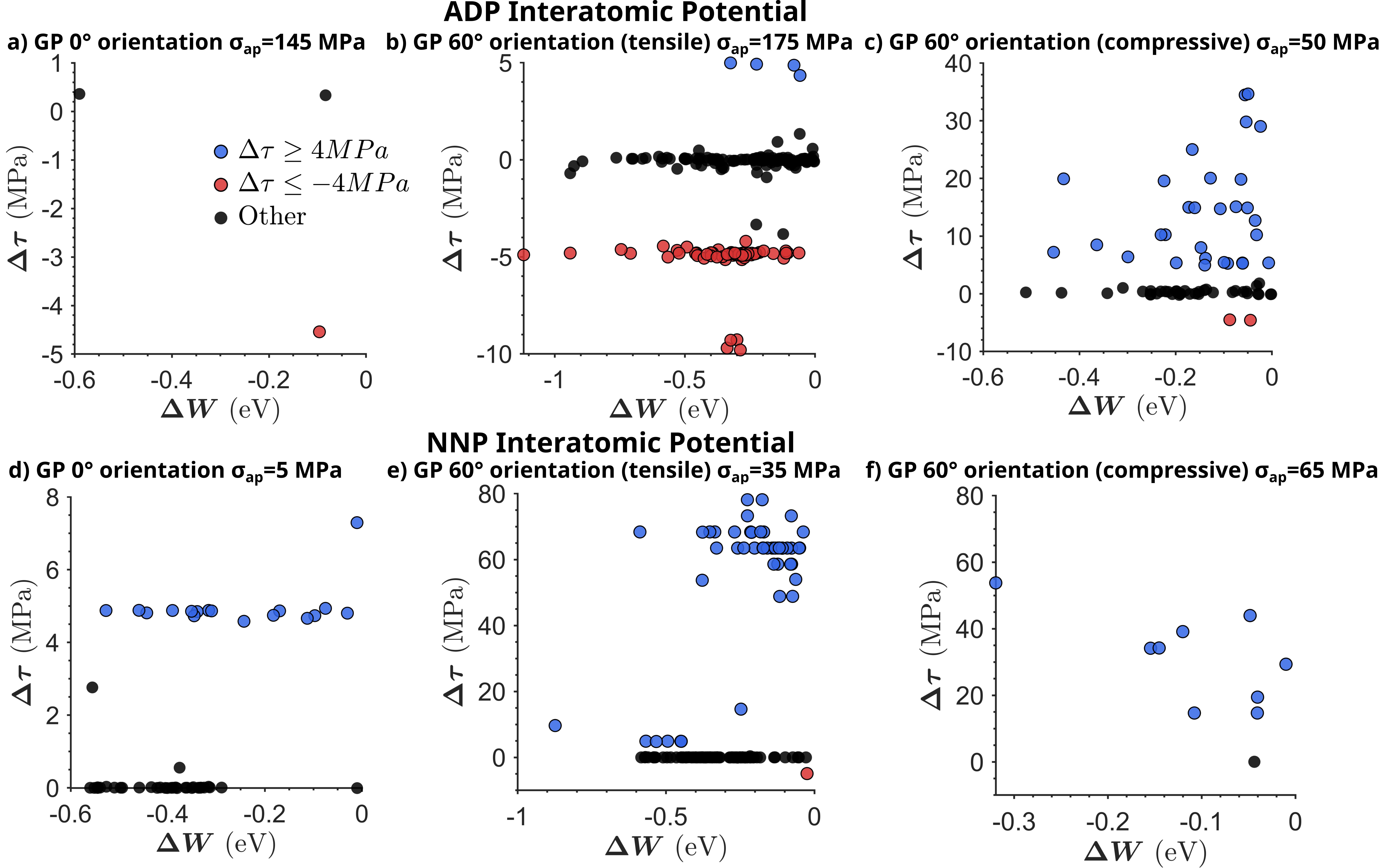}
\caption{Correlation between energy change ($\Delta W_{ij}$) and strength change ($\Delta\tau_{ij}$) for 
Cu$\leftrightarrow$Al exchanges in the GP zone. Top row: ADP potential; bottom row: NNP potential. Blue: 
strengthening ($\Delta\tau_{ij}\ge 4$~MPa); red: weakening ($\Delta\tau_{ij}\le -4$~MPa); black: neutral.}
\label{fig4}
\end{figure}

As a first step towards assessing the effect of Cu$\leftrightarrow$Al exchanges on mechanical properties, the change in resolved shear stress $\Delta\tau_{ij}$, associated with each energetically favorable exchange $\Delta W_{ij}>0$ is evaluated for both the ADP and NNP11 potentials, as shown in Fig.~\ref{fig4}. 

The results indicate that values of $\Delta W_{ij}$ and $\Delta\tau_{ij}$ are not strongly correlated for either potential. Thus, the energetic favourability of a local Cu$\leftrightarrow$Al exchange does not by itself predict the mechanical consequence of an exchange. NNP11 predicts systematically larger $\Delta\tau_{ij}$ values than ADP, with favourable exchanges yielding $\Delta\tau_{ij} \approx -5$–70~MPa, compared to -10–35~MPa with ADP. 

Fig.~\ref{fig4} shows that energetically favorable Cu$\leftrightarrow$Al exchanges can result in both strengthening and weakening for both ADP and NNP11 potentials, with the magnitude of strengthening reaching up to $\sim$80MPa and weakening limited to approximately $\sim$10MPa. For both potentials, the mechanical effect is more substantial for the GP 60$^\circ$ orientation than the GP 0$^\circ$ orientation. For the ADP potential, weakening-type exchanges are rare in the GP $0^\circ$ and GP $60^\circ$ compressive configurations, with only a single weakening event observed for GP $0^\circ$ and two weakening events for the GP $60^\circ$ compressive case, where the dislocation overcomes the GP zone via looping. In contrast, the GP $60^\circ$ tensile configuration, where cutting occurs, exhibits dozens of mechanically weakening exchange events, alongside only four mechanically strengthening exchanges. By contrast, for the NNP11 potential, where all GP zone configurations are overcome by cutting, weakening-type exchanges are essentially absent across all orientations. Energetically favorable exchanges are instead either mechanically neutral or strengthening.

\begin{figure}[!h]
\centering
\includegraphics[width=0.95\textwidth]{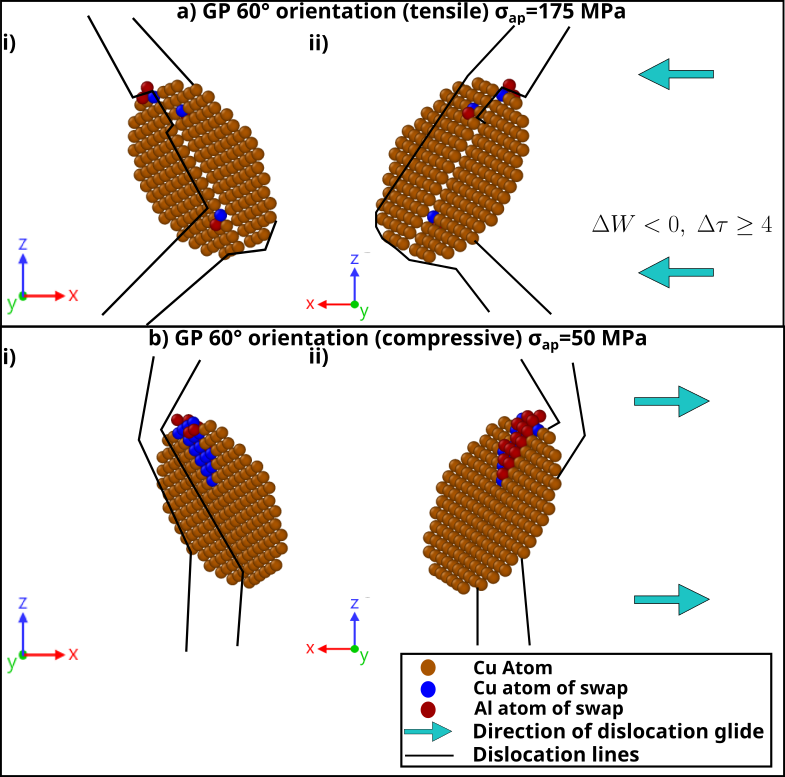}
\caption{Strengthening-type favourable exchanges ($\Delta W_{ij}<0$ and $\Delta\tau_{ij}\ge4$~MPa) for the 60$^\circ$ orientation under tensile (a) and compressive (b) loading. Brown spheres: Cu in GP zone; blue spheres: Cu atoms undergoing exchanges; red spheres: corresponding Al atoms. Black lines indicate the dislocation cores. Panels~(i) and~(ii) correspond to views taken along the $+y$ and $-y$ directions, respectively, providing complementary perspectives of the same exchange event on opposite sides of the GP zone. The green arrows show the direction of dislocation glide.}
\label{fig3}
\end{figure}

To provide further insight, the spatial location of strengthening exchanges for the ADP potential has been examined. 
By contrasting strengthening and weakening exchange events for the GP $60^\circ$ tensile configuration (Figs.~\ref{fig2}a, \ref{fig3}a, and \ref{fig:1A}), a distinction emerges among strengthening and weakening cases. Weakening exchanges preferentially cluster along the cutting path of the dislocation through the GP zone, forming extended, cut-plane-aligned rearrangements. In contrast, strengthening exchanges remain spatially localized,isolated and confined to narrow regions aligned with the dislocation, occurring primarily at the initial points of contact between the leading Shockley partial and the GP zone, near the zone boundaries. These locations correspond to the entry and exit points of the leading partial during cutting, where local shear stresses are highest. For the looping in the compressive configuration of GP\,60$^\circ$ (Fig.~\ref{fig3}b), strengthening exchanges cluster only within the compressive lobe of the GP zone, near its boundaries, and in the vicinity of the leading partial. 
\newpage
\subsection{Dynamic Strain Aging Model}

The atomistic simulations presented here reveal that energetically favorable Cu$\leftrightarrow$Al exchanges can occur when a dislocation interacts with a GP zone in Al–Cu alloys. Among the configurations examined, the GP $60^\circ$ compressive orientation exhibits the highest number of exchange events with a substantial mechanical impact. Therefore, the GP $60^\circ$ compressive orientation is the primary focus of the present section.

While the atomistic calculations resolve the energetics and mechanical impact of individual exchange events ($\Delta W_{ij}$ and $\Delta \tau_{ij}$), they do not provide direct information on the timescales over which such rearrangements occur or how their cumulative effect evolves with strain rate. To address this and quantify how strengthening develops dynamically during a dislocation–precipitate interaction, we employ a kinetic Monte Carlo (kMC) model that explicitly tracks the temporal evolution of exchange events under applied shear stress.

\begin{figure}[!h]
\centering
\includegraphics[width=0.9\textwidth]{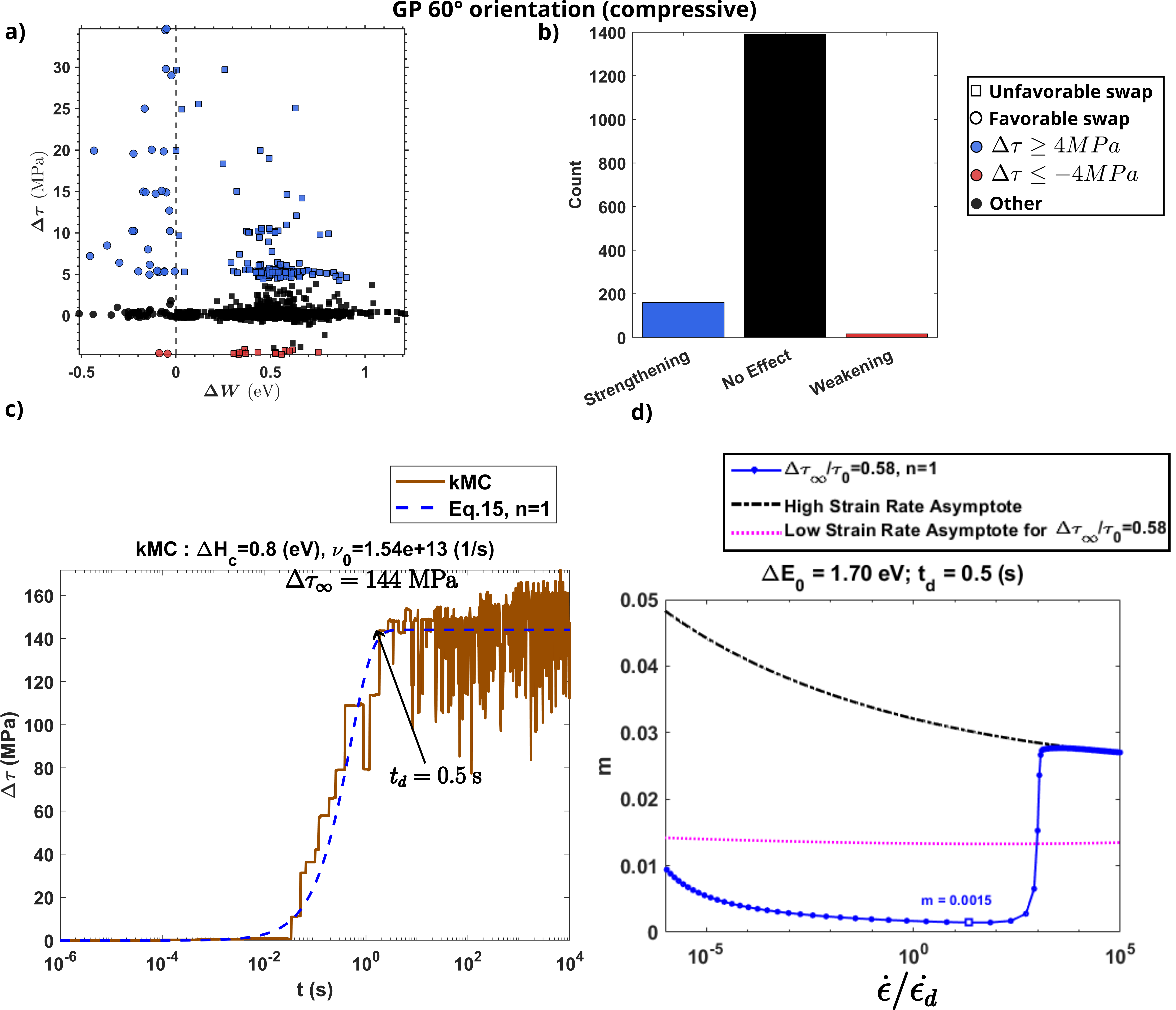}

\caption{
ADP, GP $60^\circ$, compressive loading at $\sigma_{\mathrm{ap}} = 50\,\text{MPa}$. All Cu$\leftrightarrow$Al single-hop exchanges are enumerated and classified as forward ($\Delta W_{ij} < 0$) or backward ($\Delta W_{ij} \ge 0$). (a) Change in alloy strength, $\Delta\tau_{ij}$, as a function of exchange energy difference, $\Delta W_{ij}$, with colors indicating strengthening ($\Delta\tau_{ij} \ge 4\,\text{MPa}$), weakening ($\Delta\tau_{ij} \le -4\,\text{MPa}$), and neutral events. Square markers denote backward exchanges, while circles denote forward exchanges. (b) Fractions of exchanges that lead to strengthening, weakening, or no effect.  (c) Temporal evolution of cumulative strengthening computed using a kinetic Monte Carlo (kMC) simulation of correlated exchanges, assuming an attempt frequency $\nu_0$ and $\Delta H_c = \Delta E_0 = 0.8$. (d) SRS evolution as a function of normalized strain rate predicted using the Soare--Curtin model of dynamic strain aging, using saturation strengthening and aging time from (c). 
}
\label{fig6}
\end{figure}

The kMC model samples the exchange energies $\Delta W_{ij}$ from a static dislocation stress field including both forward  and backward  Cu$\leftrightarrow$Al exchanges to capture the net evolution of strengthening during a single event of a dislocation overcoming the GP zone. As shown in Fig.~\ref{fig6}a–b, while most exchanges produce negligible mechanical change, a substantial fraction of favorable exchanges lead to a significant increases in precipitate strength $\Delta \tau_{ij}$. 

Within the kMC framework, each exchange is treated as a single-hop, independent event, forming a Markovian process in which only site occupancy history is retained. The exchanges are assumed to be uncorrelated and additive, such that the cumulative strengthening reflects the temporal accumulation of individual strengthening events. Fig.~\ref{fig6}c shows the temporal evolution of the cumulative strengthening predicted by the kMC simulations for the GP $60^\circ$ configuration. The strengthening increases rapidly at early times and then saturates, indicating the presence of a characteristic aging timescale.  This kMC simulated strengthening vs time data is fitted using the aging law in Eq.~\eqref{eq:aging}, from which two aging parameters are extracted: the saturation strengthening, $\Delta\tau_{\infty}$, and the characteristic saturation time, $t_{\mathrm{sat}}$.

The aging law Eq.~\eqref{eq:aging} is fitted using the kinetic exponent $n=1$, corresponding to barrier-limited, localized solute migration processes such as cross-core diffusion~\cite{curtin2006predictive}. This choice reproduces the general evolution of the aging-induced strengthening also referred as the back-stress $\Delta tau_a(t)$) by \cite{soare2008single}, with a sharp transition and  saturation matching the kMC data.  In the following analysis, a saturation strengthening of $\Delta\tau_{\infty} = 144$~MPa and a characteristic saturation time of $t_{\mathrm{sat}} = 0.5$\~s were used.

These parameters are subsequently incorporated into the Soare–Curtin dynamic strain aging framework~\cite{soare2008single}, which was constructed to model FCC metals in regimes dominated by a single rate-dependent strengthening mechanism. The dislocation–precipitate interaction defines the underlying energy barrier, which evolves during the pinning time due to local Cu rearrangements near the GP zone. This evolution is incorporated as a time-dependent stress shift referred to as aging-induced back-stress that modifies the effective driving stress acting on the dislocation. Using the athermal CRSS $\tau_0 = 248.67$~MPa obtained from atomistic calculations in section~\ref{sec:crss} and a strengthening ratio $\Delta\tau_{\infty}/\tau_0 = 0.58$, the model predicts substantial aging potential far exceeding the 5–15 MPa typically associated with solid-solution DSA in Al alloys~\cite{curtin2006predictive}.

The resulting SRS predicted by the Soare--Curtin DSA model is shown in Fig.~\ref{fig6}d. We first interpret this plot with respect to the high and low strain rate asymptotes. The high strain rate asymptote corresponds to the case where no diffusive rearrangement (exchanges) occurs, the energy barrier is constant in time with $\Delta F(\tau) = \Delta F_0(\tau)$ with $\Delta\tau_a(t)=0$ , and $m$ is determined by classical thermal activation theory following equations~\eqref{eq:orowan}, \eqref{eq:tw}, \eqref{eq:rate}, \eqref{eq:barrier}, \eqref{eq:aging}. The low strain rate asymptote corresponds to the case where the effect of diffusive rearrangement has saturated from the start of the simulation, the energy barrier is constant in time with $\Delta\tau_a(t)=\inf$, and $m$ is again determined by equations~\eqref{eq:orowan}, \eqref{eq:tw}, \eqref{eq:rate}, \eqref{eq:barrier}, \eqref{eq:aging}. The key takeaway from comparing these two asymptotes is that diffusive alloy atom rearrangement changes the value of $m$ over a broad range of technologically relevant strain rate\footnote{We note that with classical thermal activation models and no diffusive rearrangement (equations~\eqref{eq:orowan}, \eqref{eq:tw}, \eqref{eq:rate}, \eqref{eq:barrier}, \eqref{eq:aging}), $m$ is a nonmonotonic function of strain rate and thus an increasing energy barrier should not always be expected to lower the value of $m$ }. 

In the case of the evolving energy barrier, the value of $m$ is considerably lower than that of both asymptotes across a broad range of strain rate.  An intuitive understanding of this result can be obtained by considering a dislocation successively overcoming identical precipitates via thermal activation. Considering each dislocation-precipitate interaction to be thermally activated event, there will always be a nonzero fraction of thermally activated events where the energy barrier will have evolved to its saturated state. In the intermediate strain rate regime with a low value of $m$, a small change in stress will dramatically change the fraction of these long timescale events, dramatically change the average time per thermally activated event, and hence dramatically change the macroscopic strain rate. From the inverse perspective, a small change in strain rate will have a minuscule effect on the stress, hence a low value of $m$. Alternatively, at high or low rates outside this intermediate regime, the strain rate is effectively determined by the thermal activation of barriers in either their initial or saturated state, and hence corresponds to the asymptotic cases mentioned in the previous paragraph.  

Importantly, this outcome is insensitive to the form and parameters of the model. For instance, a change in the saturation strength from 144 MPa to 25MPa, would change $m$ in the intermediate regime from $ m \approx 1.5\times 10^{-3}$ to $ m \approx 5\times 10^{-3}$. Moreover, a regime of low $m$ would appear even in the case where diffusive rearrangement leads to a net weakening, as it is the heterogeneity in obstacle activation energy that develops that results in the region of low $m$. Supportively, studies on static heterogeneous obstacle populations have been shown to also drastically reduce $m$ relative to homogeneous obstacles~\cite{picu2009strain}. 

Quantitatively, experimentally measured strain rate sensitivities of precipitate-strengthened aluminum alloys are 
$m\approx0.005$~\cite{byrne1961precipitate,muraishi2002strengthening, khan2012variable,gupta2017strain}. These values are much more similar to the values of $m$ found in Fig.~\ref{fig6}d than those of previous dislocation-precipitate atomistic and continuum modeling which did not allow for diffusive at alloy rearrangement (nearest-neighbor exchanges), 
$m\approx0.04$,~\cite{saroukhani2016harnessing,singh2013atomistic,xu2007thermally}. Nonetheless, it is important to emphasize that the SRS reported here is not necessarily governing the strain-rate sensitivity of a bulk material, as the heterogeneity of the initial precipitate population and the correlated behavior of dislocation ensembles can also control $m$ values~\cite{picu2009strain,mcdowell2024nonequilibrium1,mcdowell2024nonequilibrium2,mcdowell2024nonequilibrium3,mcdowell2025nonequilibrium4}. 

Strain-rate sensitivity is often used as an indicator of the rate-controlling deformation mechanism. In particular, departures from the ordinarily low $m$ values of structural alloys are commonly interpreted as evidence of a mechanism change, for example toward creep, superplasticity, or grain-boundary-mediated deformation at low strain rates, or toward dynamic dislocation effects at high strain rates~\cite{mukherjee1968experimental,chichili1998high,wei2004effect,gianola2006increased}. The results of this section suggest that such interpretations should be made with caution, because $m$ can increase substantially at both low and high strain rates even in the absence of a change in the underlying mechanism. More broadly, the results of this section may help explain the scatter in reported SRS values for alloys of similar composition~\cite{picu2010aluminum}.

\section{Conclusion}
Consistent with expectation, our atomistic modeling indicates that isolated monolayer GP zones are stable against nearest-neighbour Cu$\leftrightarrow$Al exchanges. The modeling further indicates that when a dislocation is pinned at a GP zone, nearest-neighbour Cu$\leftrightarrow$Al exchanges are likely to occur considering available knowledge on the kinetics of such atom exchanges. 

In contrast to the literature on deformation induced precipitate dissolution and reconstruction~\cite{behnood1989plastic,brechet1996pseudo,zhu1998strain,ovri2015new,rowlands2023portevin,zhang2026interplay}, the redistribution of alloy elements presented here is local, involving only nearest neighbor atom exchange events as opposed to the longer range transport of alloy atoms. We find that the energetically favored local exchange of alloy atoms leads to an overall strengthening of the precipitate-dislocation junction. We believe that these two findings are unlikely to be an artifact of the interatomic potential, as both have been arrived at with two independent interatomic potentials of different form, ADP and NNP11.

By integrating the full atomistic exchange catalog into a kinetic Monte Carlo model, we obtained a time-dependent strengthening from which we extracted the saturation value of strengthening at room temperature and the onset time of that saturation. Incorporating the lower-scale parameters into the Soare--Curtin DSA framework provides a mechanistic link between diffusive alloy atom rearrangements at dislocation-precipitate junctions and the rate dependence of plastic flow, $m$, in the absence of higher scale dislocation avalanche type behavior~\cite{picu2009strain, mcdowell2024nonequilibrium1,mcdowell2024nonequilibrium2,mcdowell2024nonequilibrium3,mcdowell2025nonequilibrium4}. The predicted value of $m$ is near zero over a broad range of strain rate, distinctly different from previous studies of dislocation-precipitate interactions in the absence of alloy redistribution, but consistent with experimental behavior of age-hardened Al--Cu alloys. 
\section{Acknowledgments}
The primary support of this project was by the U.S. National Science Foundation grant No.~1922081. Support from the US Office of Naval Research (N629092512049) and the NZ Ministry of Business, Innovation, and Employment, New Zealand (UOC2454) are also gratefully acknowledged.
\newpage
\appendix
\section{Supplementary Information}
\subsection{Strengthening- and weakening-type favorable exchanges for the $60^\circ$ orientation under tensile loading (ADP potential)}
\begin{figure}[H]
\centering
\includegraphics[width=0.99\textwidth]{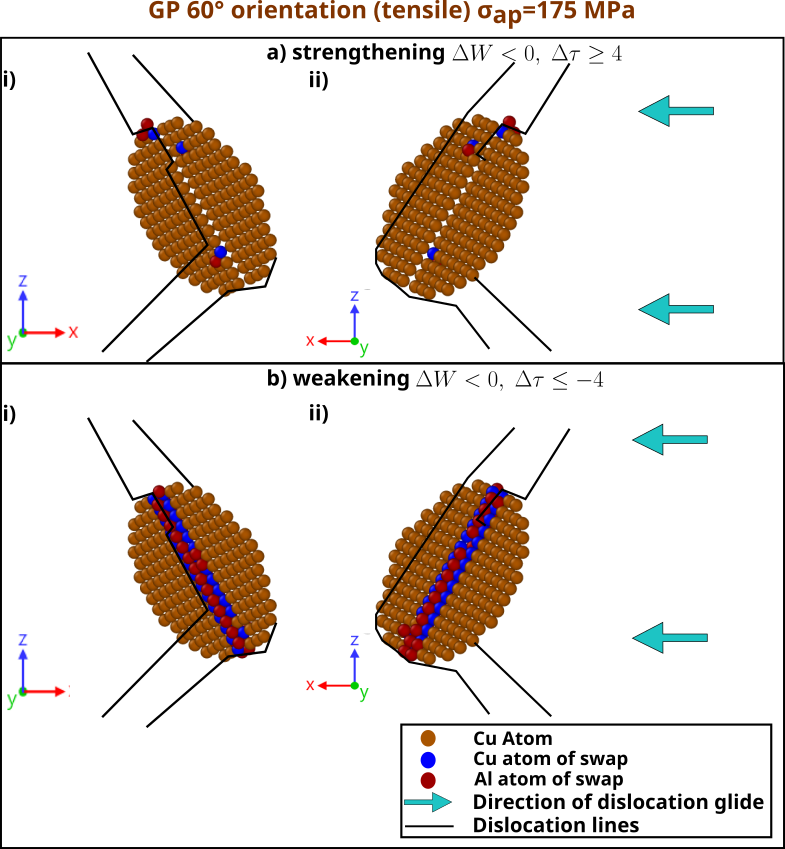}
\caption{The 60$^\circ$ orientation under tensile loading with both a) strengthening ($\Delta W_{ij}<0$ and $\Delta\tau_{ij}\ge4$~MPa) and b) weakening ($\Delta W_{ij}<0$ and $\Delta\tau_{ij}\le=4$~MPa) favourable exchanges highlighted. }
\label{fig:1A}
\end{figure}

In this section, Fig.~\ref{fig:1A} illustrates Cu$\leftrightarrow$Al exchange events in the GP $60^\circ$ tensile configuration under an applied shear stress of $\sigma_{\mathrm{ap}} = 175$~MPa, classified according to their mechanical consequence focusing on only weakening and strengthening favorable exchanges. Panels~(a.i--ii) show mechanically strengthening exchanges ($\Delta W < 0$, $\Delta \tau \ge 4$~MPa), while panels~(b.i--ii) show mechanically weakening exchanges ($\Delta W < 0$, $\Delta \tau \le -4$~MPa). Among energetically favorable exchanges that have a measurable mechanical impact on the strength of the GP zone, only a small subset evolve into strengthening configurations, whereas the majority form spatially correlated rearrangements along the cutting path that promote mechanical weakening. This trend indicates that alloy atom rearrangements activated by the cutting of the GP zone are, on average, more likely to reduce the local resistance to dislocation motion than to enhance it.

\subsection{Strengthening-type favorable exchanges for all GP orientations using the NNP potential}
\begin{figure}[H]
\centering
\includegraphics[width=0.8\textwidth]{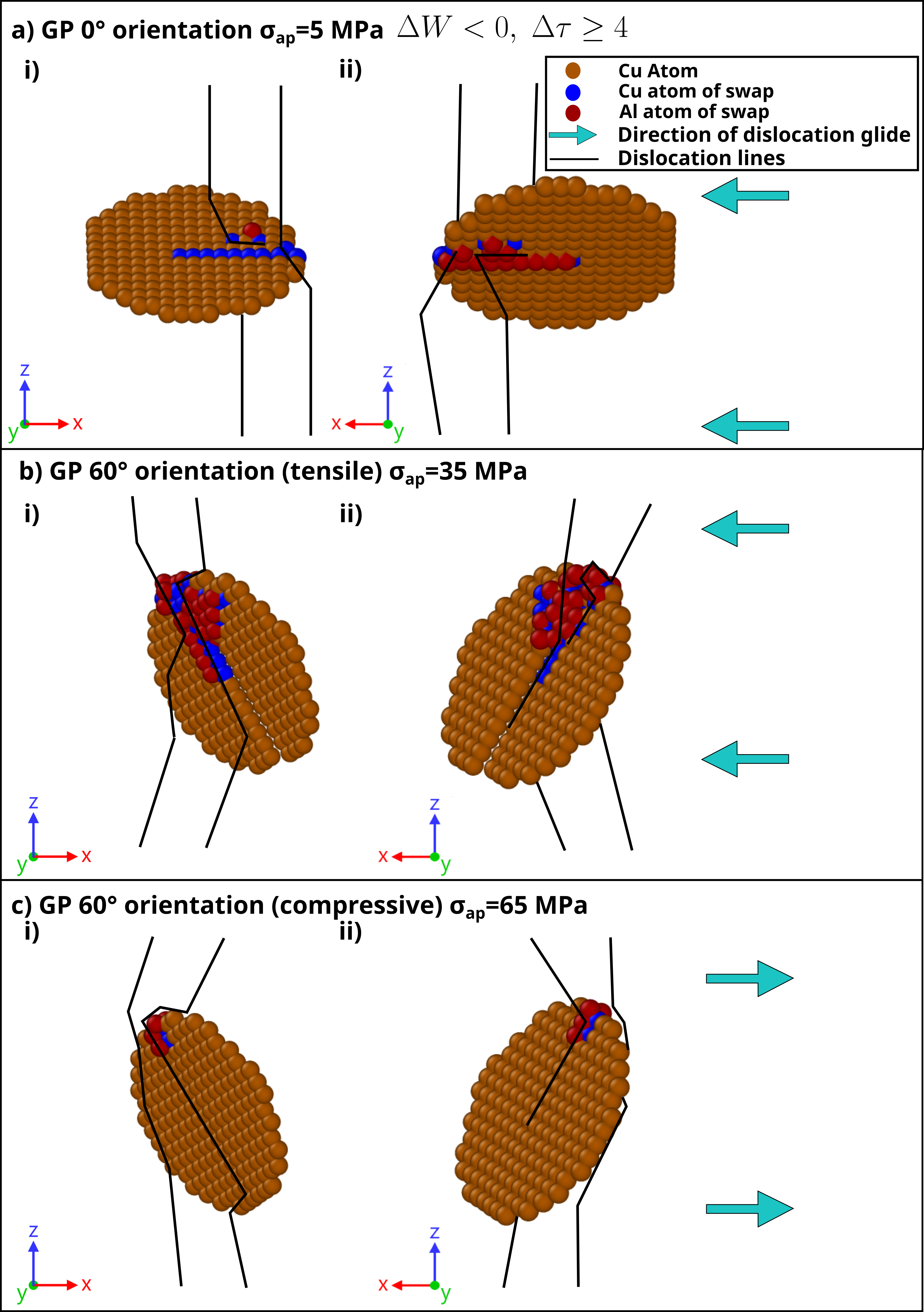}
\caption{Strengthening-type favorable Cu--Al exchange events 
($\Delta W < 0$, $\Delta \tau \ge 4$~MPa) for GP zones interacting with an edge dislocation, shown for different GP orientations and loading conditions using the NNP potential.}
\label{fig:2A}
\end{figure}

Fig.~\ref{fig:2A} illustrates representative strengthening-type favorable Cu--Al exchange events 
($\Delta W < 0$, $\Delta \tau \ge 4$~MPa) occurring when an edge dislocation is pinned at a GP zone modeled with the NNP potential. For the GP $0^\circ$ orientation, the dislocation partially cut the GP zone, whereas for the GP $60^\circ$ tensile configuration, the GP zone is fully cut. In contrast, the GP $60^\circ$ compressive configuration remains largely unchanged, and is not cut.  Across all cases, the strengthening-type exchanges are spatially clustered along the cutting path, but are not distributed along its entire length; instead, they are localized near the region where the leading partial initiates cutting of the GP zone. For the GP $60^\circ$ tensile case, compared to the GP $0^\circ$, the favorable strengthening exchanges look less like a long correlated line and more like a local “patch” of rearrangement at/near the dislocation–precipitate intersection still aligned with the cutting line .

\bibliographystyle{unsrt} 
\bibliography{references} 
\end{document}